\documentclass[aps,prc,twocolumn,superscriptaddress]{revtex4-1}

\usepackage{graphicx}
\usepackage{amsmath}
\usepackage{braket}
\usepackage{url}
\usepackage{multirow}
\usepackage{multirow}
\usepackage{amssymb}
\usepackage{booktabs}
\usepackage{rotating}
\usepackage{bm}
\usepackage{bbm}
\usepackage{ulem}
\usepackage{makecell}
\usepackage{mathrsfs}
\usepackage{booktabs}
\usepackage{xcolor}

\usepackage[colorlinks,
linkcolor=blue,
anchorcolor=blue,
urlcolor=blue,
citecolor=blue]{hyperref}

\usepackage[mathlines]{lineno}

\begin{document}


\newcommand{\rwYQ}[1]{ {\color{green} #1}}
\newcommand{\rwFSQ}[1]{{\color{blue}#1}}
\newcommand{\rFSQ}[1]{{\color{cyan}#1}}




\title{First \textit{ab initio} calculations of first-forbidden $\beta$ transitions in the reactor antineutrino anomaly}


\author{X. Y. Xu}
 \affiliation{%
School of Physics, and State Key Laboratory of Nuclear Physics and Technology, Peking University, Beijing 100871, China
}%
\author{Z. Y. Meng}
 \affiliation{%
School of Physics, and State Key Laboratory of Nuclear Physics and Technology, Peking University, Beijing 100871, China
}%

\author{Z. C. Xu}
\affiliation{Key Laboratory of Nuclear Physics and Ion-beam Application (MOE), Institute of Modern Physics, Fudan University, Shanghai 200433, China}
\affiliation{Shanghai Research Center for Theoretical Nuclear Physics, NSFC and Fudan University, Shanghai 200438, China}

\author{F. R. Xu}
 \email{frxu@pku.edu.cn}
\affiliation{%
School of Physics, and State Key Laboratory of Nuclear Physics and Technology, Peking University, Beijing 100871, China
}
\affiliation{Southern Center for Nuclear-Science Theory, Institute of Modern Physics, Chinese Academy of Sciences, Huizhou 516000, China}


\begin{abstract}
Forbidden $\beta$ transitions are important for understanding the reactor antineutrino anomaly. Starting from chiral two- plus three-nucleon forces, we have derived the valence-space effective Hamiltonian and effective operators of first-forbidden transitions using the many-body perturbation theory. 20 dominant first-forbidden transitions have been investigated, which provide important contributors to the reactor antineutrino spectrum anomaly.  Calculated $\log ft$ values are in reasonable agreement with experimental data. Obtained shape factors exhibit significant deviations from the values approximated with forbidden transitions treated as allowed transitions.
The ``5 MeV bump'' observed in the experimental ${}^{235}$U-fission antineutrino spectrum was discussed with self-consistent shape factors obtained in the present \textit{ab initio} calculations. Unlike phenomenological models that require empirical quenching factors to reproduce $\beta$-decay data, the present \textit{ab initio} calculations do not need to introduce quenching factors for calculations of the first-forbidden transitions. 
\end{abstract}{}
\maketitle
\section{\label{sec:level1}Introduction}
Nuclear $\beta$ decay has played an important role in nuclear physics by offering a precise window into the weak interaction and its possible extensions beyond the Standard Model~\cite{RevModPhys.78.991, RevModPhys.87.1483}. The allowed $\beta$ transition has long been under investigation~\cite{PhysRevLett.10.531,Prog.Theor.Phys.49.652, PhysRevC.91.025501, P.Gysbers_nature.15.428(2019),Phys.Lett.B.832.137259, PhysRevC.108.L031301,PhysRevC.111.034309, PhysRevLett.133.211801, annurev-nucl-2024-020726}. The forbidden $\beta$ transition has also attracted growing interest due to the important roles in understanding rapid neutron-capture process in cosmic nucleosynthesis~\cite{PhysRevC.87.025803, PhysLettB.756.2016, PhysRevC.97.054321}, in searching for new physics~\cite{PhysLettB.285.2017, PhysRevLett.134.081805} and in interpreting the reactor antineutrino ($\bar{\nu}_e$) flux~\cite{PhysRevLett.112.202501, PhysRevC.91.025503, PhysRevC.99.031301, PhysRevC.108.055501, Prog.Part.Nucl.Phys.136.104106, PhysRevLett.124.232502}. However, theoretical studies of forbidden $\beta$ transitions remain scarce, and {\it ab initio} calculations are particularly limited~\cite{PhysRevC.110.014324}.

\textit{Ab initio} calculations, based on nuclear forces derived from the chiral effective field theory (EFT)~\cite{E.Epelbaum_RevModPhys.81.1773(2009), R.Machleidt_Phys.Rep.503.1(2011)}, have made great progresses~\cite{CC_PhysRevC.82.034330(2010), SCGF_PhysRevC.87.011303(2013),IMSRG-Phys.Rep.621.165(2016),MRIMSRG_PhysRevC.87.034307(2013), bognerNonperturbativeShellModelInteractions2014, S.R.Stroberg_PhysRevLett.118.032502(2017), S.R.Stroberg_PhysRevLett.126.022501(2021),GIMSRG_PhysRevC.99.061302(2019),DIMSRG_PhysRevC.105.L061303(2022),XuNST2024,LiNST2024}, and become a cornerstone of state-of-the-art nuclear theoretical calculations. Several \textit{ab initio} methods, such as the many-body perturbation theory (MBPT)~\cite{MBPT14PRC,MBPTBSHU,tichaiHartreeFockManybody2016} and in-medium similarity renormalization group (IMSRG)~\cite{firstIMSRG2011, IMSRG-Phys.Rep.621.165(2016), strobergNonempiricalInteractionsNuclear2019}, have been widely used to derive valence-space effective interactions by decoupling a small valence space from the full large Hilbert space, thereby incorporating the influence of the excluded space without introducing empirical parameters. Within the same consistent renormalization and decoupling framework, one can also derive other effective operators (e.g., relevant to electroweak observables), and the same realistic nuclear interaction is the sole input.

The reactor antineutrino flux can be obtained theoretically via conversion of reactor electron spectra into antineutrino, for which the HM method~\cite{Prog.Part.Nucl.Phys.136.104106, PhysRevC.83.054615, PhysRevC.84.024617} has served as a standard reference since 2011. However, the HM model predicts a reactor antineutrino 
flux approximately 6\% higher than the experimental data measured in short-baseline reactor experiments~\cite{PhysRevD.83.073006}. This discrepancy became known as the reactor antineutrino anomaly (RAA) and was later corroborated by the newest generation of inverse $\beta$ decay (IBD) experiments (Daya Bay~\cite{PhysRevLett.134.201802},
RENO~\cite{PhysRevD.104.L111301}, and Double Chooz~\cite{Nat.Phys.16.558}) carried out at commercial reactor facilities. In addition to the rate anomaly, these experiments also observed a robust spectral distortion with a bump relative to the HM prediction in the 4-7 MeV energy region, far beyond the model uncertainty, and is commonly referred to as the ``5 MeV bump''~\cite{PhysRevLett.116.211801, PhysRevD.104.L111301, PhysRevLett.134.201802}.

One widely discussed explanation of the RAA invokes a sterile neutrino with a mass of the order of 1 eV (or higher), which could cause antineutrino 
disappearance at very short baselines (a few meters) via additional neutrino oscillations~\cite{Prog.Part.Nucl.Phys.111.103736}. To date, however, no direct experimental evidence has confirmed the existence of such a state. 
We found that in the HM model the forbidden transitions are treated using highly simplified approximations~\cite{PhysRevC.83.054615, PhysRevC.84.024617}. Forbidden transitions, particularly first-forbidden transitions, should be better treated. They contribute significantly to the total spectrum, and dominate the spectrum in the energy higher than 4 MeV~\cite{Prog.Part.Nucl.Phys.136.104106}. The highly simplified approximations can introduce substantial uncertainties~\cite{PhysRevLett.112.202501, PhysRevC.91.025503, PhysRevC.99.031301, PhysRevD.100.053005} in the predictions of the antineutrino flux and spectrum. Of course, precise calculations of forbidden transitions are challenging, as they depend sensitively on nuclear structure details and on the interplay of multipole transition operators, and are difficult to be treated in a fully self-consistent manner.

In the previous works~\cite{PhysRevC.91.025503, PhysRevC.99.031301, PhysRevC.100.054323, PLB2024138515}, first-forbidden $\beta$ transitions were studied in a self-consistent manner but using phenomenological Hamiltonian and bare transition operators with empirical quenching factors needed. In this letter, starting from EFT nuclear forces, we perform the first \textit{ab initio} calculations of first-forbidden transitions using effective Hamiltonian and effective operators obtained via the MBPT, without the need for empirical quenching factors. 

\section{\label{sec:level2}The Formalism}
We start from the intrinsic $A$-body nuclear Hamiltonian,
\begin{equation}
H= \sum_{i=1}^A\left(1-\frac{1}{A}\right) \frac{\boldsymbol{p}_i^2}{2 m}+\sum_{i<j}^A\left(v_{i j}^{\mathrm{NN}}-\frac{\boldsymbol{p}_i \cdot \boldsymbol{p}_j}{m A}\right) +\sum_{i<j<k}^A v_{i j k}^{\mathrm{3N}},
\label{H_in}
\end{equation}
where $\boldsymbol{p}_i$ is the nucleon momentum in the laboratory system, and $m$ is the nucleon mass, while $v^{\mathrm{NN}}$ and  $v^{\mathrm{3N}}$ are the two-nucleon force (2NF) and three-nucleon force (3NF), respectively. Hamiltonian~(\ref{H_in}) is transferred to the Hartree-Fock basis, and rewritten in the normal-ordered form of operators as follows
\begin{equation}
\begin{split}
    H=&E_0+\sum_{ij}h_{ij}:a^{\dagger}_ia_j:+\frac{1}{2!^2}\sum_{ijkl}V_{ijkl}:a^{\dagger}_ia^{\dagger}_ja_la_k:\\
    &+\frac{1}{3!^2}\sum_{ijklmn}\mathcal{W}_{ijklmn}:a^{\dagger}_ia^{\dagger}_ja^{\dagger}_ka_na_ma_l:,
\end{split}
\label{H_norm}
\end{equation}
where $E_0$, $h$, $V$, and $\mathcal{W}$ are normal-ordered zero-, one-, two-, and three-body terms, respectively. In the normal-ordering process, 3NF effects are absorbed into the normal-ordered zero-, one- and two-body terms. It is usually accepted that 3NF effects can be well captured at the normal-ordered two-body (NO2B) level~\cite{PhysRevLett.107.072501,R.Roth_PhysRevLett.109.052501(2012)}, while the residual normal-ordered three-body term $\mathcal{W}$ can be neglected in practical calculations. With this normal-ordered Hamiltonian, we employ the MBPT to construct the valence-space effective Hamiltonian and other effective operators by the so-called $\hat{Q}$-box and $\hat{\Theta}$-box folded diagrams defined in Refs.~\cite{PhysRevC.110.024308, Fan_2026}. With the effective Hamiltonian and $\beta$-transition operators obtained by MBPT, we can evaluate $\beta$ decays of nuclei.

The partial half life $t$ for a $\beta$ transition from the initial state to the final state is expressed in terms of the phase-space factor $f$ as 
\begin{equation}
f t=K,
\end{equation}
where $K=6147 \mathrm{~s}$ is determined by some fundamental constants~\cite{Nucl.Phys.A.509.429-460_1990_hardy}. In this letter, we restrict ourselves to the $\beta^-$ decay, which is the case for the RAA. Natural units are adopted in this letter. The phase-space factor has the following form,
\begin{equation}
f=\int_1^{W_0} C(W) F_0(Z, W)\left(W^2-1\right)^{1 / 2} W\left(W_0-W\right)^2 d W,
\label{phase factor}
\end{equation}
where $W$ is the total energy of the emitted electron, expressed in the unit of the electron mass, and $W_0$ is the endpoint energy of the $\beta$ spectrum. $C(W)$ denotes the shape factor, and $F_0(Z,W)$ is the well-known Fermi function with $Z$ being the proton number. The shape factor $C(W)$ can be written as~\cite{BEHRENS1971111NPA, Behrens1982}

\begin{equation}
\begin{aligned}
C\left(W\right)= & \sum_{k_e, k_{\bar{\nu}}, K} \lambda_{k_e}\left[M_K\left(k_e, k_{\bar{\nu}}\right)^2+m_K\left(k_e, k_{\bar{\nu}}\right)^2\right. \\
& \left.-\frac{2 \gamma_{k_e}}{k_e w_e} M_K\left(k_e, k_{\bar{\nu}}\right) m_K\left(k_e, k_{\bar{\nu}}\right)\right],
\end{aligned}
\label{C(W) shape factor}
\end{equation}
where the integers $k_e$ and $k_{\bar{\nu}}$ are related to the angular momentum quantum numbers of the partial-wave expansion of the electron and antineutrino wave functions, respectively. The index $K$ labels the tensor rank of the forbidden $\beta$-decay operator, ranging from 0 to 2 for the first-forbidden transitions. $\gamma_{k_e}=\sqrt{k_e^2-(\alpha Z)^2}$ is an auxiliary quantity with the fine-structure constant $\alpha=1/137$. $\lambda_{k_e}$ is defined as~\cite{Behrens1982}
\begin{equation}
\lambda_{k_e}=\frac{F_{k_e-1}\left(Z, W\right)}{F_0\left(Z, W\right)},
\end{equation}
where $F_{k_e-1}(Z,W)$ is the generalized Fermi function defined as 
\begin{equation}
\begin{aligned}
F_{k_e-1}\left(Z, W\right)= & 4^{k_e-1}\left(2 k_e\right)\left(k_e+\gamma_{k_e}\right)\left[\left(2 k_e-1\right)!!\right]^2 e^{\pi y} \\
& \times\left(\frac{2 p R}{\hbar}\right)^{2\left(\gamma_{k_e}-k_e\right)}\left(\frac{\left|\Gamma\left(\gamma_{k_e}+i y\right)\right|}{\Gamma\left(1+2 \gamma_{k_e}\right)}\right)^2,
\end{aligned}
\end{equation}
where $p$ is the electron momentum, and $R$ is the nuclear radius taken from experimental data~\cite{DATATABLEANGELI201369}. $\Gamma$ denotes the Gamma function (see~\cite{Iddrisu_Tetteh_Mathsci_2017}) with  $y=\alpha ZW/p$. The nuclear-structure information is encoded in the quantities $M_K(k_e, k_{\bar{\nu}})$ and $m_K(k_e, k_{\bar{\nu}})$, which involve complicated combinations of different nuclear matrix elements and leptonic phase-space factors. For first-forbidden transitions, $k_e+ k_{\bar{\nu}}$ can take the values 2 or 3. According to Ref.~\cite{BEHRENS1971111NPA}, for first-forbidden transitions, $M_K(k_e, k_{\bar{\nu}})$ and $m_K(k_e, k_{\bar{\nu}})$ are defined as
\begin{subequations}
    \begin{align}
M_0(1,1) =& {}^{\mathrm{A}}F_{000}^{0}
 - \frac{1}{3}\alpha Z\,{}^{\mathrm{A}}F_{011}^{0}(1,1,1,1) \nonumber
 \\
 &- \frac{1}{3}W_{0}R\,{}^{\mathrm{A}}F_{011}^{0},
\\
M_1(1,1) =& -{}^{\mathrm{V}}F_{101}^{0}
 - \frac{1}{3}\alpha Z\sqrt{\frac{1}{3}}\,{}^{\mathrm{V}}F_{110}^{0}(1,1,1,1) \nonumber
 \\
 &- \frac{1}{3}W_{0}R\sqrt{\frac{1}{3}}\,{}^{\mathrm{V}}F_{110}^{0}
- \frac{1}{3}(W-q)R\sqrt{\frac{2}{3}}\,{}^{\mathrm{A}}F_{111}^{0}\nonumber
 \\
 &- \frac{1}{3}\alpha Z\sqrt{\frac{2}{3}}\,{}^{\mathrm{A}}F_{111}^{0}(1,1,1,1),
\\
M_1(1,2) =& \frac{1}{3}qR\!\left[
 \sqrt{\frac{2}{3}}\,{}^{\mathrm{V}}F_{110}^{0}
 + \sqrt{\frac{1}{3}}\,{}^{\mathrm{A}}F_{111}^{0}
\right],
\\
M_1(2,1) =& \frac{1}{3}pR\!\left[
 \sqrt{\frac{2}{3}}\,{}^{\mathrm{V}}F_{110}^{0}
 - \sqrt{\frac{1}{3}}\,{}^{\mathrm{A}}F_{111}^{0}
\right],
\\
M_2(1,2) =& -\frac{1}{3}qR\,{}^{\mathrm{A}}F_{211}^{0},
\\
M_2(2,1) =& -\frac{1}{3}pR\,{}^{\mathrm{A}}F_{211}^{0},
\\
m_0(1,1) =& -\frac{1}{3}R\,{}^{\mathrm{A}}F_{011}^{0},
\\
m_1(1,1) =& -\frac{1}{3}R\!\left[
 \sqrt{\frac{1}{3}}\,{}^{\mathrm{V}}F_{110}^{0}
 + \sqrt{\frac{2}{3}}\,{}^{\mathrm{A}}F_{111}^{0}
\right],
\end{align}
\end{subequations}
where $q$ denotes the antineutrino momentum, and the form factor coefficient ${}^{\mathrm{V/A}}F_{KLS}^{0}$ is labeled by the vector (V) or axial-vector (A) nature of the weak current with $L$ and $S$ denoting the orbital and spin ranks of the transition operator, respectively. The superscript 0 indicates the leading term retained in the impulse approximation~\cite{BEHRENS1971111NPA}. For first-forbidden transitions, the form factor ${}^{\mathrm{V/A}}F_{KLS}^{0}$ can be expressed as~\cite{BEHRENS1971111NPA,PhysRevC.87.025803}
\begin{subequations}
\begin{align}
{}^{\mathrm{A}}F_{011}^{0}
&= g_A\sqrt{3}\,
\frac{\bigl\langle f\bigl\|\sum_k
\bigl[\boldsymbol{C}_{1}^{k}\times\boldsymbol{\sigma}^{k}\bigr]^{0}
\,\boldsymbol{t}_{-}^{k}\bigr\| i\bigr\rangle}{\sqrt{2J_i+1}},
\\
{}^{\mathrm{V}}F_{110}^{0}
&= \sqrt{3}\,
\frac{\bigl\langle f\bigl\|\sum_k
\boldsymbol{C}_{1}^{k}\,\boldsymbol{t}_{-}^{k}\bigr\| i\bigr\rangle}{\sqrt{2J_i+1}},
\\
{}^{\mathrm{A}}F_{111}^{0}
&= g_A\sqrt{3}\,
\frac{\bigl\langle f\bigl\|\sum_k
\bigl[\boldsymbol{C}_{1}^{k}\times\boldsymbol{\sigma}^{k}\bigr]^{1}
\,\boldsymbol{t}_{-}^{k}\bigr\| i\bigr\rangle}{\sqrt{2J_i+1}},
\\
{}^{\mathrm{A}}F_{211}^{0}
&= g_A\sqrt{3}\,
\frac{\bigl\langle f\bigl\|\sum_k
\bigl[\boldsymbol{C}_{1}^{k}\times\boldsymbol{\sigma}^{k}\bigr]^{2}
\,\boldsymbol{t}_{-}^{k}\bigr\| i\bigr\rangle}{\sqrt{2J_i+1}},
\\[2pt]
{}^{\mathrm{A}}F_{011}^{0}(1,1,1,1)
&= g_A\sqrt{3} \nonumber
\\
&\quad\times
\frac{\bigl\langle f\bigl\|\sum_k
I(1,1,1,1)\,
\bigl[\boldsymbol{C}_{1}^{k}\times\boldsymbol{\sigma}^{k}\bigr]^{0}
\,\boldsymbol{t}_{-}^{k}\bigr\| i\bigr\rangle}{\sqrt{2J_i+1}},
\\
{}^{\mathrm{V}}F_{110}^{0}(1,1,1,1)
&= \sqrt{3}\,
\frac{\bigl\langle f\bigl\|\sum_k
I(1,1,1,1)\,\boldsymbol{C}_{1}^{k}\,\boldsymbol{t}_{-}^{k}\bigr\| i\bigr\rangle}{\sqrt{2J_i+1}},
\\
{}^{\mathrm{A}}F_{111}^{0}(1,1,1,1)
&= g_A\sqrt{3} \nonumber
\\
&\quad\times
\frac{\bigl\langle f\bigl\|\sum_k
I(1,1,1,1)\,
\bigl[\boldsymbol{C}_{1}^{k}\times\boldsymbol{\sigma}^{k}\bigr]^{1}
\,\boldsymbol{t}_{-}^{k}\bigr\| i\bigr\rangle}{\sqrt{2J_i+1}},
\\[2pt]
{}^{\mathrm{A}}F_{000}^{0}
&= \frac{g_A\sqrt{3}}{m}\,
\frac{\bigl\langle f\bigl\|\sum_k
\bigl[\boldsymbol{\sigma}^{k}\times\nabla^{k}\bigr]^{0}
\,\boldsymbol{t}_{-}^{k}\bigr\| i\bigr\rangle}{\sqrt{2J_i+1}},
\\
{}^{\mathrm{V}}F_{101}^{0}
&= -\frac{1}{m}\,
\frac{\bigl\langle f\bigl\|\sum_k
\nabla^{k}\,\boldsymbol{t}_{-}^{k}\bigr\| i\bigr\rangle}{\sqrt{2J_i+1}},
\end{align}
\end{subequations}
where
\begin{equation}
\boldsymbol{C}_{l l_m}=\sqrt{\frac{4 \pi}{2 l+1}} \boldsymbol{Y}_{l l_m},
\end{equation}
with $\boldsymbol{Y}_{l l_m}$ denoting the spherical harmonics. Here, $\boldsymbol{\sigma}$ is the Pauli spin operator, and $\boldsymbol{t}_{-}$ is the isospin lowering operator. The labels $i$ and $f$ refer to the initial and final nuclear states, respectively. $I(1,1,1,1)$ is a Coulomb factor depending on the nuclear charge distribution, which can be found in Ref.~\cite{BEHRENS1971111NPA}. The weak axial coupling constant takes $g_A=-1.2701$~\cite{PhysRevD.86.010001}.

\section{\label{sec:level3}Calculations and discussions} 
In this letter, we have calculated 20 dominant first-forbidden transitions 
which govern the reactor antineutrino
spectrum in prompt energy higher than 4 MeV, and account for more than half of the total forbidden-transition probability~\cite{PhysRevC.91.011301,DataSheetsBROWN20181}. They also contribute about 40$\%$ of the total cumulative flux in the energy region between 4 and 7 MeV~\cite{PhysRevC.108.055501,PhysRevC.91.011301}. In the present calculations, we take the valence space consisting of $\pi\{0 f_{5/2}$, $1 p_{3/2}$, $1 p_{1/2}$, $0 g_{9/2}\}$ and $\nu\{0g_{7/2}$, $1d_{5/2}$, $1d_{3/2}$, $2s_{1/2}\}$ which reaches our maximal limit of the feasible SM diagonalizing.

The newly developed chiral EFT potential labeled by $\mathrm{NN + 3N}$(lnl)~\cite{P.Gysbers_nature.15.428(2019)} was used in this letter. This interaction takes a 2NF truncated at N$^{4}$LO level and 3NF at N$^{2}$LO. 3NF was produced using both local and nonlocal (lnl) regulators~\cite{P.Gysbers_nature.15.428(2019), somaNovelChiralHamiltonian2020}. The 2NF and 3NF are consistently evolved using the similarity renormalization group (SRG) at a momentum resolution scale $\lambda = 2.0~\mathrm{fm}^{-1}$. With the same nuclear forces, we first performed the Hartree-Fock (HF) calculation to obtain the HF basis which is used in the evolution of the effective Hamiltonian (also effective operators). The HF calculation starts within a harmonic-oscillator (HO) basis at the optimized
frequency $\hbar \Omega=16~\mathrm{MeV}$~\cite{P.Gysbers_nature.15.428(2019)} with 15 major shells (i.e., $e= \left.2 n+l \leqslant e_{\mathrm{max} }=14\right)$ and $E_{3\mathrm{max} }=e_1+e_2+e_3 \leqslant 24$ for 3NF, which is sufficient to get calculations
converged~\cite{binderInitioPathHeavy2014,miyagiConvergedInitioCalculations2022}. Then, the MBPT is processed in the self-consistent Hartree-Fock basis~\cite{PhysRevC.110.024308} to obtain the valence-space effective Hamiltonian. This Hamiltonian is diagonalized using the large-scale SM code, KSHELL~\cite{kshell_Comput.Phys.Commun.244.372(2019)}.

Our discussions begin by presenting the calculated log$ft$ values, along with the corresponding experimental data. Experimental $\beta$-decay $Q$ values~\cite{DataSheetsBROWN20181, ensdf} were used  to minimize errors arising from $Q$ value. It is worth noting that $Q$ values calculated by MBPT with the present EFT forces do not differ much from the experimental values. Table~\ref{logft_J_0} shows the log$ft$ values of the dominant first-forbidden transitions for the RAA with $\Delta J=0$ (i.e., zero spin change between the initial and final states). 
Overall, the calculated results reproduce reasonably the experimental $\log ft$ values without introducing any quenching factor which is often required in phenomenological calculations~\cite{PhysRevC.87.025803, PhysRevC.97.054321}. Although the theoretical values are slightly higher than the data, the deviations remain within an acceptable range.
\begin{table}[ht]
\renewcommand\arraystretch{1.3}
    \caption{Calculated \label{logft_J_0}log$ft$ values of the $\Delta J=0$ dominant first-forbidden transitions for the RAA, compared with data~\cite{DataSheetsBROWN20181, ensdf}.}
 \begin{ruledtabular}
  \begin{tabular}{ cccccc}
\multicolumn{2}{c}{Transitions} &
\multicolumn{4}{c}{log$ft$}
\\
\cline{1-2}
\cline{3-6}
   Initial&Final&Expt.&MBPT
   \\	
   \hline
  \multicolumn{1}{l}{$^{89}$Br($3/2_1^{-}$)}  & \multicolumn{1}{c}{$^{89}$Kr($3/2_1^{+}$)}&
\multicolumn{1}{c}{---}  
&
\multicolumn{1}{c}{7.13}     
\\
  \multicolumn{1}{l}{$^{90}$Rb($0_1^-$)}  & \multicolumn{1}{c}{$^{90}$Sr($0_1^+$)}&
\multicolumn{1}{c}{7.81}  
&
\multicolumn{1}{c}{7.85}     
\\
  \multicolumn{1}{l}{$^{91}$Kr($5/2_1^{+}$)}  & \multicolumn{1}{c}{$^{91}$Rb($5/2_1^{-}$)}&
\multicolumn{1}{c}{6.36}  
&
\multicolumn{1}{c}{6.74}      
\\
 \multicolumn{1}{l}{$^{92}$Rb($0_1^-$)}  & \multicolumn{1}{c}{$^{92}$Sr($0_1^+$)}&
\multicolumn{1}{c}{5.75}  
&
\multicolumn{1}{c}{6.20}     
\\
 \multicolumn{1}{l}{$^{93}$Rb($5/2_1^{-}$)}  & \multicolumn{1}{c}{$^{93}$Sr($5/2_1^{+}$)}&
\multicolumn{1}{c}{6.14}  
&
\multicolumn{1}{c}{6.37}      
\\
 \multicolumn{1}{l}{$^{94}$Y($2_1^-$)}  & \multicolumn{1}{c}{$^{94}$Zr($2_1^+$)}&
\multicolumn{1}{c}{7.18}  
&
\multicolumn{1}{c}{7.20}      
\\
 \multicolumn{1}{l}{$^{95}$Rb($5/2_1^{-}$)}  & \multicolumn{1}{c}{$^{95}$Sr($5/2_1^{+}$)}&
\multicolumn{1}{c}{6.01}  
&
\multicolumn{1}{c}{6.86}     
\\
 \multicolumn{1}{l}{$^{95}$Sr($1/2_1^{+}$)}  & \multicolumn{1}{c}{$^{95}$Y($1/2_1^{-}$)}&
\multicolumn{1}{c}{6.16}  
&
\multicolumn{1}{c}{7.31}     
\\
 \multicolumn{1}{l}{$^{96}$Y($0_1^-$)}  & \multicolumn{1}{c}{$^{96}$Zr($0_1^+$)}&
\multicolumn{1}{c}{5.59}  
&
\multicolumn{1}{c}{6.32}      
\\
 \multicolumn{1}{l}{$^{97}$Y($1/2_1^{-}$)}  & \multicolumn{1}{c}{$^{97}$Zr($1/2_1^{+}$)}&
\multicolumn{1}{c}{5.70}  
&
\multicolumn{1}{c}{6.81}      
\\
  \end{tabular}
 \end{ruledtabular}
\end{table}

Table~\ref{logft_J_1} shows the log$ft$ values with $\Delta J=1$. In general, the theoretical results agree with the experimental data without introducing any quenching factors, though deviations from data are larger than 10\% in $^{87}$Se($3/2_1^{+}$), $^{89}$Br($3/2_1^{-}$) and $^{95}$Rb($5/2_1^{-}$) transitions. Such discrepancies are also commonly seen in \textit{ab initio} calculations of allowed transitions~\cite{P.Gysbers_nature.15.428(2019)}. It is mainly attributed to the neglect of the coupling of the weak force to two nucleons~\cite{P.Gysbers_nature.15.428(2019)}, known as two-body currents (2BCs). However, formulating 2BCs for forbidden transitions is highly complex, leaving current phenomenological treatments limited. The meson-exchange-current enhancement has been only applied to the axial charge matrix element ${}^{\mathrm{A}}F_{000}^{0}$~\cite{KOSTENSALO2018480, PhysRevC.100.054323}, while 2BC corrections to the remaining first-forbidden operators remain unexplored. A complete \textit{ab initio} treatment of 2BCs in forbidden transitions would be the focus of our future work. Similar discrepancies are also observed for the $\Delta J=2$ unique transitions in Table~\ref{logft_J_2}, where the theoretical results are systematically lower than the experimental values. For $\Delta J=0$ and 1 transitions, 2BC corrections to multiple contributing operators may partially cancel one another, mitigating their net effect, whereas for $\Delta J=2$ transitions only ${}^A\!F_{211}^0$ contributes at leading order and no such cancellation is possible, potentially making the net 2BC effect more pronounced. Nevertheless, given that forbidden transitions are sensitive to nuclear structure details, the overall agreement supports the reliability of the present MBPT calculations. 

\begin{table}[ht]
\renewcommand\arraystretch{1.3}
    \caption{Calculated \label{logft_J_1}log$ft$ values of the $\Delta J=1$ dominant first-forbidden transitions for the RAA, compared with data~\cite{DataSheetsBROWN20181, ensdf}.}
 \begin{ruledtabular}
  \begin{tabular}{ cccc}
\multicolumn{2}{c}{Transitions} &
\multicolumn{2}{c}{log$ft$}
\\
\cline{1-2}
\cline{3-4}
   Initial&Final&Expt.&MBPT
   \\	
   \hline
  \multicolumn{1}{l}{$^{86}$Br($1_1^{-}$)}  & \multicolumn{1}{c}{$^{86}$Kr($0_1^{+}$)}&
\multicolumn{1}{c}{7.45}  
&
\multicolumn{1}{c}{7.16}     
\\
  \multicolumn{1}{l}{$^{86}$Br($1_1^-$)}  & \multicolumn{1}{c}{$^{86}$Kr($2_1^+$)}&
\multicolumn{1}{c}{7.72}  
&
\multicolumn{1}{c}{7.41}     
\\
  \multicolumn{1}{l}{$^{87}$Se($3/2_1^{+}$)}  & \multicolumn{1}{c}{$^{87}$Br($5/2_1^{-}$)}&
\multicolumn{1}{c}{6.10}  
&
\multicolumn{1}{c}{7.35}    
\\
  \multicolumn{1}{l}{$^{89}$Br($3/2_1^{-}$)}  & \multicolumn{1}{c}{$^{89}$Kr($5/2_1^{+}$)}&
\multicolumn{1}{c}{6.50}  
&
\multicolumn{1}{c}{7.54}     
\\
  \multicolumn{1}{l}{$^{91}$Kr($5/2_1^{+}$)}  & \multicolumn{1}{c}{$^{91}$Rb($3/2_1^{-}$)}&
\multicolumn{1}{c}{6.69}  
&
\multicolumn{1}{c}{6.95}    
\\
  \multicolumn{1}{l}{$^{95}$Rb($5/2_1^{-}$)}  & \multicolumn{1}{c}{$^{95}$Sr($7/2_1^{+}$)}&
\multicolumn{1}{c}{6.03}  
&
\multicolumn{1}{c}{7.57}    
\\
  \multicolumn{1}{l}{$^{95}$Rb($5/2_1^{-}$)}  & \multicolumn{1}{c}{$^{95}$Sr($3/2_1^{+}$)}&
\multicolumn{1}{c}{---}  
&
\multicolumn{1}{c}{7.76}     
\\
  \end{tabular}
 \end{ruledtabular}
\end{table}

\begin{table}[ht]
\renewcommand\arraystretch{1.3}
    \caption{\label{logft_J_2} Calculated log$ft$ values of the $\Delta J=2$ dominant first-forbidden transitions for the RAA, compared with data~\cite{DataSheetsBROWN20181, ensdf}.}
 \begin{ruledtabular}
  \begin{tabular}{ cccccc}
\multicolumn{2}{c}{Transitions} &
\multicolumn{2}{c}{log$ft$}
\\
\cline{1-2}
\cline{3-4}
   Initial&Final&Expt.&MBPT
   \\	
   \hline
\multicolumn{1}{l}{$^{88}$Rb($2_1^{-}$)}  & \multicolumn{1}{c}{$^{88}$Sr($0_1^{+}$)}&
\multicolumn{1}{c}{9.25}  
&
\multicolumn{1}{c}{7.13}     
\\
  \multicolumn{1}{l}{$^{94}$Y($2_1^{-}$)}  & \multicolumn{1}{c}{$^{94}$Zr($0_1^{+}$)}&
\multicolumn{1}{c}{9.35}  
&
\multicolumn{1}{c}{7.42}    
\\
  \multicolumn{1}{l}{$^{95}$Rb($5/2_1^-$)}  & \multicolumn{1}{c}{$^{95}$Sr($1/2_1^+$)}&
\multicolumn{1}{c}{$\geqslant 10.2$}  
&
\multicolumn{1}{c}{10.56}    
\\
  \end{tabular}
 \end{ruledtabular}
\end{table}

\begin{figure}[ht]
    \includegraphics[width=0.48\textwidth]{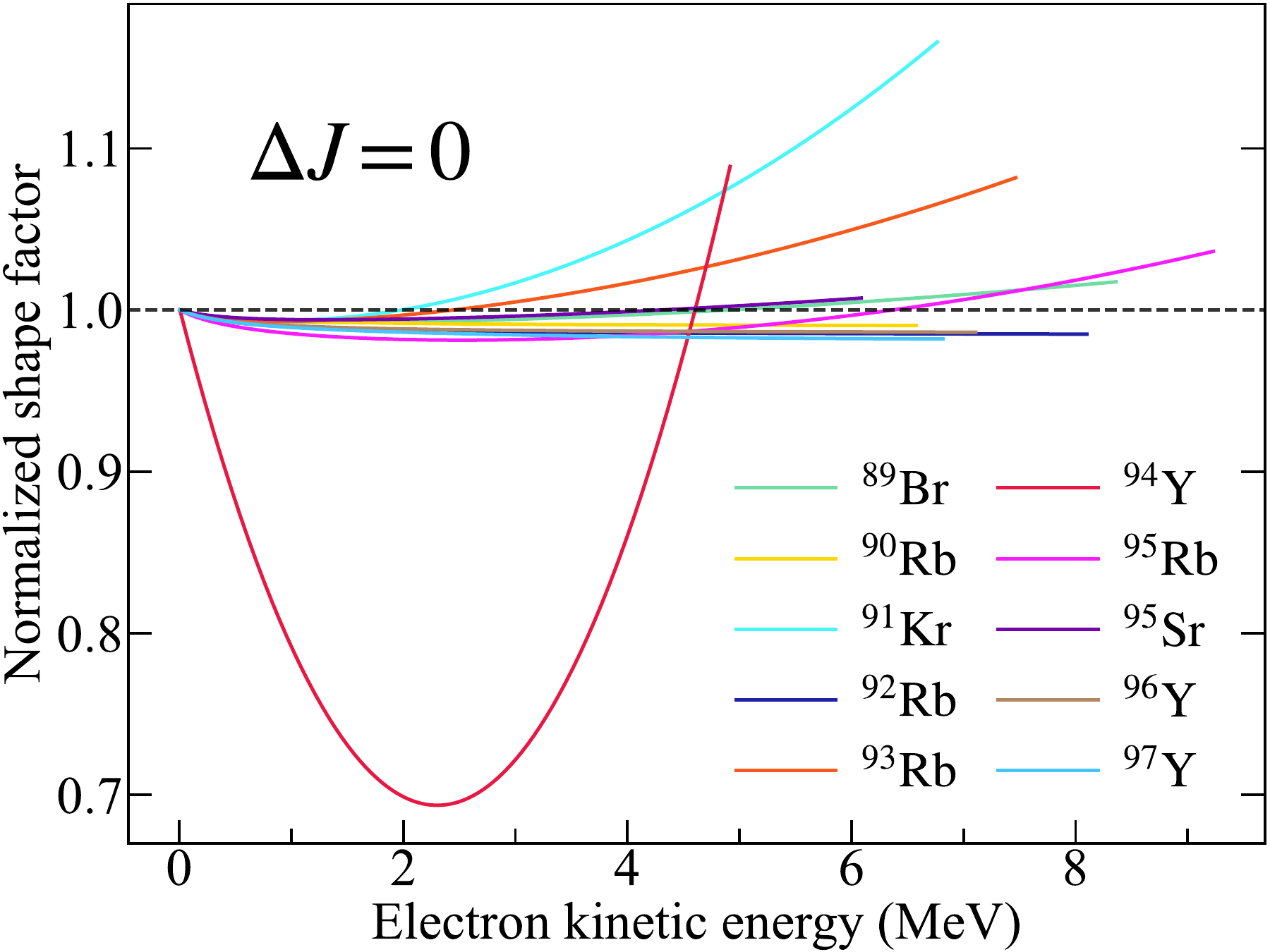}
    \caption{\label{J_0_shape_factor}Calculated normalized shape factor for $\Delta J=0$ first-forbidden transitions as a function of the kinetic energy of the emitted electron. For allowed transitions, the normalized shape factor is unity, indicated by the black dashed line.} 
\end{figure}

The nuclear structure information is embedded in the shape factor expressed by Eq.~(\ref{C(W) shape factor}). Figure~\ref{J_0_shape_factor} illustrates the calculated normalized shape factor for $\Delta J=0$ first-forbidden transitions as a function of electron kinetic energy. As mentioned already, forbidden transitions in reactor fission products are usually approximated as allowed transitions in conventional models~\cite{PhysRevC.84.024617}. For allowed transitions, the normalized shape factor is unity~\cite{Behrens1982}. However, the present \textit{ab initio} calculations show that the shape factors of first-forbidden transitions clearly deviate from unity, as shown in Fig.~\ref{J_0_shape_factor}. The shape factor of the first-forbidden transition $^{94} \mathrm{Y}\left(2_1^{-}\right) \rightarrow{ }^{94} \mathrm{Zr}\left(2_1^{+}\right)$ exhibits a parabolic curve with the electron kinetic energy (shown by the red line in Fig.~\ref{J_0_shape_factor}), which is consistent with the phenomenological calculation~\cite{PhysRevC.100.054323}. The parabolic shape arises because the $2^{+}$ excited state of $^{94}$Zr can be interpreted as a collective vibration with a wave function similar to the $0^{+}$ ground state. This results in a shape factor similar to that of the $\Delta J=2$ transition which has a parabolic shape factor as seen later.

\begin{figure}[ht]
    \includegraphics[width=0.48\textwidth]{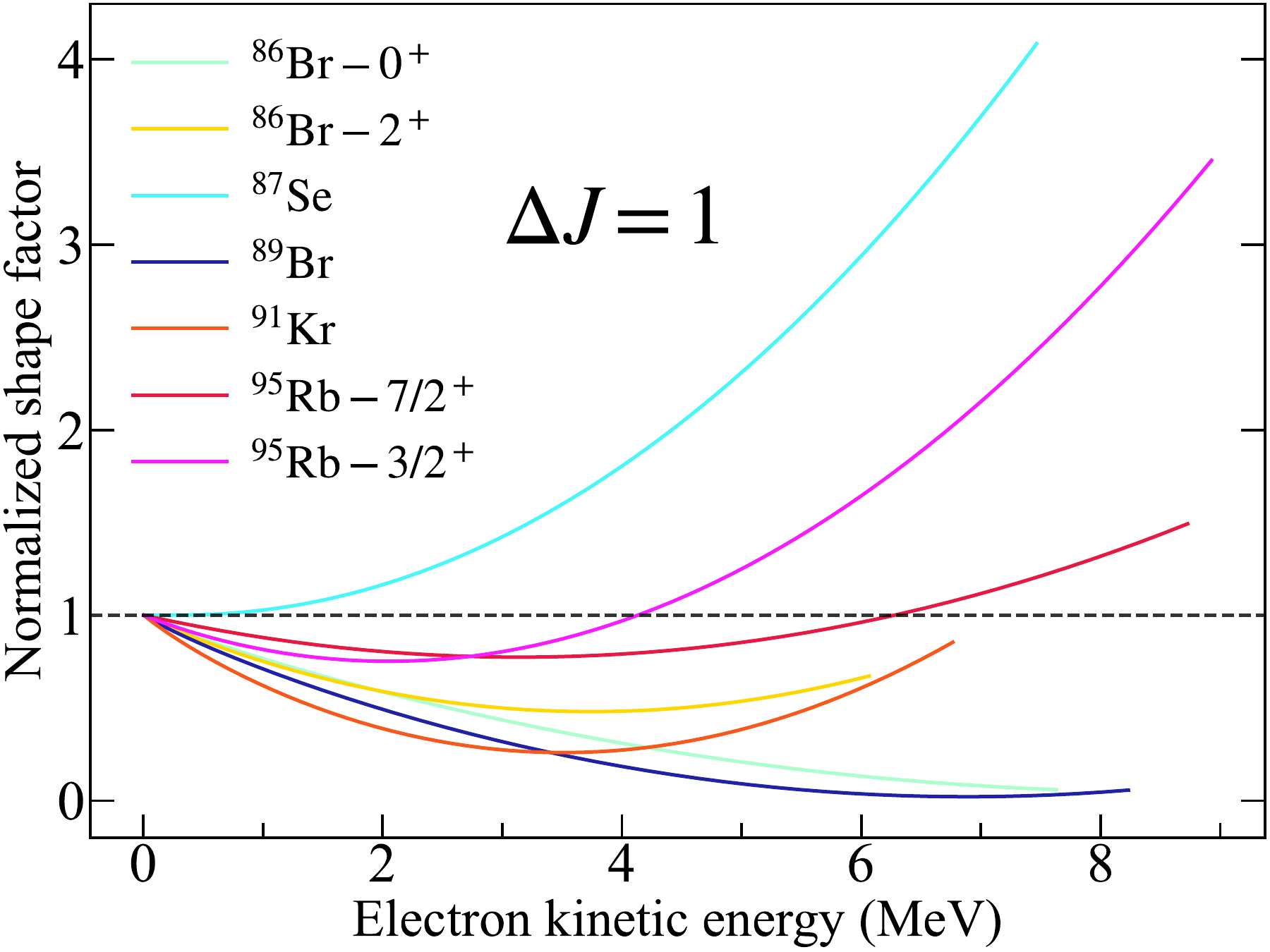}
    \caption{\label{J_1_shape_factor}Calculated normalized shape factor for $\Delta J=1$ first-forbidden transitions as a function of electron kinetic energy. Transitions to different final states are labeled by the initial nucleus and the final states of the daughter in the legend.} 
\end{figure}

\begin{figure}[ht]
    \includegraphics[width=0.48\textwidth]{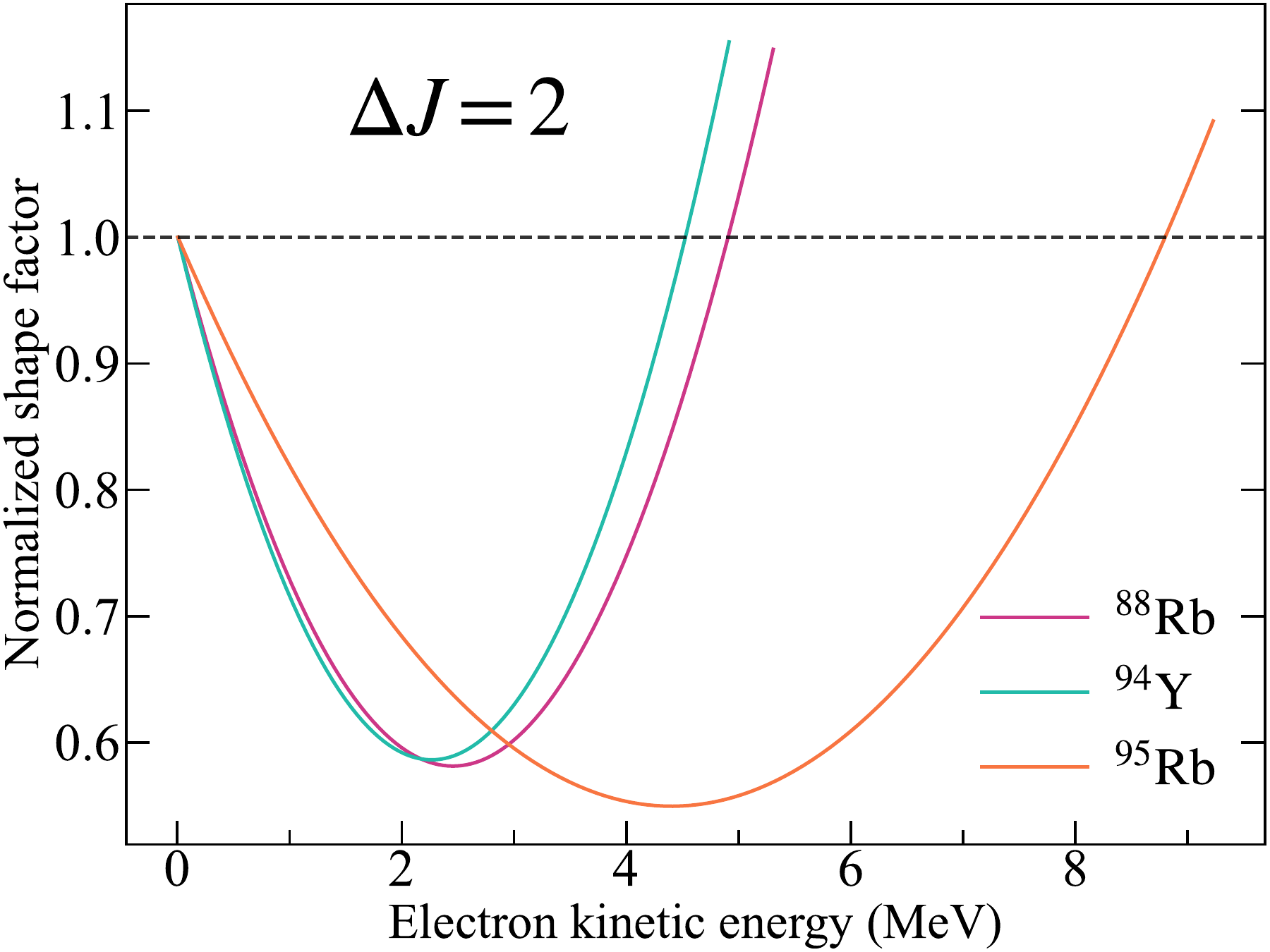}
    \caption{\label{J_2_shape_factor}Calculated normalized shape factor for $\Delta J=2$ first-forbidden transitions as a function of electron kinetic energy.} 
\end{figure}

Figures \ref{J_1_shape_factor} and \ref{J_2_shape_factor} present calculated normalized shape factors for $\Delta J=1$ and 2 first-forbidden transitions, respectively. We see that all the calculated shape factors clearly deviate from unity which is the value of an allowed transition. The $\Delta J=1$ transitions even show more pronounced deviations from unity. It is seen that the $\Delta J=2$ transitions have parabolic shape factors, which is consistent with the phenomenological calculation~\cite{PhysRevC.100.054323}. This would indicate that the difference of shape factors between allowed and forbidden transitions may not be neglected in the calculation of the energy spectrum distribution. The self-consistent calculation of forbidden transitions would be necessary in the analysis of nuclear fission processes in the reactor. 
 
The ``5 MeV bump'' remains an open question~\cite{PhysRevLett.116.211801, PhysRevD.104.L111301, PhysRevLett.134.201802}. 
The antineutrino spectrum can be evaluated by~\cite{Behrens1982}
\begin{align}
S_{\bar{\nu}}(W_{\bar{\nu}})=& C(W_0 -W_{\bar{\nu}}) F_0(Z, W_0 -W_{\bar{\nu}}) \nonumber \\
&\times[\left(W_0 -W_{\bar{\nu}})^2-1\right]^{1 / 2} (W_0 -W_{\bar{\nu}})W_{\bar{\nu}}^2,
\label{Antineutrino spectrum}
\end{align}
where $W_{\bar{\nu}}$ is the antineutrino energy. In fact, the antineutrino spectrum is exactly the integrand in Eq.~(\ref{phase factor}) but with the electron energy $W$ replaced by $W_0-W_{\bar{\nu}}$ because the total energy of the electron and antineutrino is conserved at the $\beta$-spectrum endpoint energy $W_0$ (thus $W=W_0-W_{\bar{\nu}}$). As defined in Eqs.~(\ref{phase factor}) and (\ref{C(W) shape factor}), $C(W_0 -W_{\bar{\nu}})$ and $F_0(Z, W_0 -W_{\bar{\nu}})$ are the shape factor and Fermi function, respectively.
 

In this letter, we focus on the antineutrino spectrum of ${ }^{235}$U fission products. The first-forbidden transitions calculated contribute about 40\% to the total antineutrino spectrum of the ${ }^{235}$U fission in the energy region between 4 and 7 MeV~\cite{PhysRevC.108.055501,PhysRevC.91.011301}. By summing the antineutrino spectra of the first-forbidden transitions, weighted by their respective branching ratios and cumulative fission yields of ${}^{235}$U, we obtain their total contribution to the antineutrino spectrum of the ${}^{235}$U fission. The branching ratios are taken from ENSDF~\cite{ensdf} and the cumulative fission yields of ${ }^{235}$U are obtained from the latest ENDF database~\cite{DataSheetsBROWN20181}.

\begin{figure}[ht]
    \includegraphics[width=0.48\textwidth]{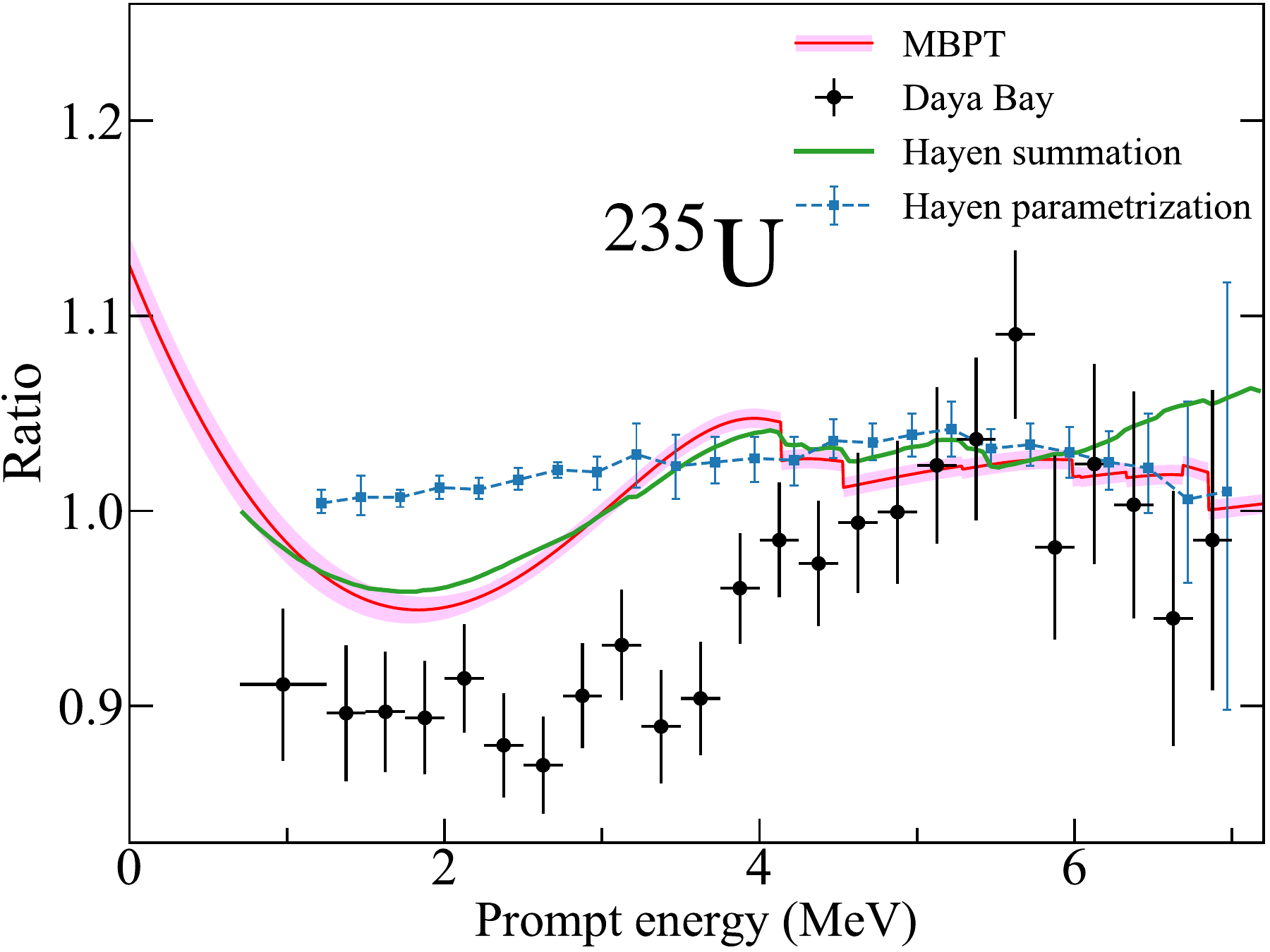}
\caption{\label{U235_spectrum}The newest Daya Bay data~\cite{PhysRevLett.134.201802} 
of the ${}^{235}$U-fission total antineutrino spectrum (ratio to the HM prediction), 
and the ratio $S^{\text{F}}_{\bar{\nu}}(\text{MBPT})/S^{\text{F}}_{\bar{\nu}}(C=1)$ 
from the present \textit{ab initio} calculation. Also shown are ``Hayen summation'' 
results obtained using shell model with 
phenomenological Hamiltonians for 29 first-forbidden transitions~\cite{PhysRevC.99.031301}, and the ``Hayen parametrization'' result incorporating parametrized 
first-forbidden shape factors for $\sim$2000 branches via Monte Carlo sampling with corresponding uncertainties~\cite{PhysRevC.100.054323}. The 
shadow indicates the theoretical uncertainty from the present \textit{ab initio} many-body calculations. 
The prompt energy is related to the antineutrino energy by $E_{\bar{\nu}}\approx 
E_{\text{prompt}}+0.782~\mathrm{MeV}$~\cite{Prog.Part.Nucl.Phys.136.104106}.}
\end{figure}

As mentioned above, in the evaluation by the HM model for the fission of ${ }^{235}$U, forbidden transitions are approximated as allowed transitions by assuming a unity shape factor of $C(W)=1$~\cite{Prog.Part.Nucl.Phys.136.104106, PhysRevC.84.024617}. However, in the present self-consistent calculations as shown in Figs.~\ref{J_0_shape_factor}-\ref{J_2_shape_factor}, the shape factors of forbidden transitions are energy-dependent (not equal to unity), which may affect the spectrum. The total experimental antineutrino spectrum is usually expressed using the ratio to a model prediction (e.g., to the HM model) by a definition as $S^\text{T}_{\bar{\nu}}(\text{Data})/S^\text{T}_{\bar{\nu}}{(\text{Model})}$ ~\cite{Prog.Part.Nucl.Phys.136.104106, PhysRevLett.116.211801, PhysRevD.104.L111301, PhysRevLett.134.201802}, where $S^\text{T}_{\bar{\nu}}$ denotes the total spectrum. In Fig.~\ref{U235_spectrum}, we plot the newest data~\cite{PhysRevLett.134.201802} of the ${ }^{235}$U-fission total antineutrino spectrum (ratio to the HM model), showing the ``5 MeV bump". The total antineutrino spectrum can contain thousands of $\beta$ decays from fission products. In the present \textit{ab initio} calculations, we are not able to calculate such large number of $\beta$ decays, but focus on the 20 dominant first-forbidden transitions mentioned above. As done in Refs.~\cite{PhysRevLett.112.202501, PhysRevC.99.031301}, we define a ratio by $S^{\text{F}}_{\bar{\nu}}(\text{MBPT})/S^{\text{F}}_{\bar{\nu}}{(C=1)}$, where $S^{\text{F}}_{\bar{\nu}}(\text{MBPT})$ denotes the total contribution from the 20 dominant first-forbidden transitions calculated using the self-consistent \textit{ab initio} MBPT shape factors, otherwise the $C=1$ approximation is applied. 

Figure~\ref{U235_spectrum} shows the calculation of $S^{\text{F}}_{\bar{\nu}}(\text{MBPT})/S^{\text{F}}_{\bar{\nu}}{(C=1)}$, with the experimental total spectrum from the Daya Bay experiment~\cite{PhysRevLett.134.201802}. We see that the measured spectrum below 4 MeV is slightly lower than the HM prediction with a ratio around 0.9. Our calculation shows a parabolic behavior of the contribution from the dominant first-forbidden transitions in energy below 4 MeV, which is mainly attributed to the $\Delta J=2$ first-forbidden transitions. 
In higher energies, the experimental spectrum exhibits a notable bump around 5 MeV.
Our calculation also shows a bump beginning at about 4 MeV. 
As pointed out in Refs.~\cite{PhysRevC.99.031301, PhysRevC.100.054323}, this bump is largely due to the downward-sloping shape factors of electrons as shown in Figs.~\ref{J_0_shape_factor}-\ref{J_2_shape_factor}. The downward slopes lead to the redistribution of the electron flux toward lower energies, whereas the antineutrino flux is toward higher energies~\cite{PhysRevC.100.054323}. 

Two phenomenological calculations given in Refs.~\cite{PhysRevC.99.031301, PhysRevC.100.054323} are also shown in Fig.~\ref{U235_spectrum} to provide comparisons with our \textit{ab initio} results. The ``Hayen summation'' result~\cite{PhysRevC.99.031301} in Fig.~\ref{U235_spectrum} treated 29 first-forbidden transitions using the nuclear shell model with phenomenological Hamiltonians, bare operators and an empirical quenched axial coupling constant $g_A = 0.9$, with an additional meson-exchange-current (MEC) enhancement factor $\epsilon_{\mathrm{MEC}} = 1.4$ applied to the axial charge matrix element ${}^{\mathrm{A}}F_{000}^{0}$. The antineutrino spectra of these transitions are then summed, weighted by their respective experimental branching ratios and cumulative fission yields of ${}^{235}$U. The ratio to the $C(W)=1$ approximation is shown in Fig.~\ref{U235_spectrum}. The ``Hayen parametrization'' result~\cite{PhysRevC.100.054323} corresponds to the Monte Carlo calculation 
incorporating the parametrized first-forbidden shape factor as $C = 1 + aW + b/W + cW^2$, which extends the treatment to $\sim$2000 first-forbidden branches in the $^{235}$U fission database. 
The uncertainty was estimated by the range of shape-factor parameterization in the Monte Carlo sampling~\cite{PhysRevC.100.054323}, which is the main uncertainty source holding about 1.5--3.0\% in the region around 5 MeV in the spectrum. Other subdominant uncertainties are from the virtual-branch conversion procedure and residuals in fitting to the Institut Laue-Langevin electron data~\cite{PhysRevC.100.054323}. The Hayen parametrization approximation does not give the "5 MeV bump".

The phenomenological Hayen summation approximation~\cite{PhysRevC.99.031301} exhibits the bump beginning around 4 MeV, which is consistent with the present \textit{ab initio} MBPT calculation, and agrees with data with the spectrum enhancement in the 4--7 MeV region, see Fig.~\ref{U235_spectrum}. 
Note that the phenomenological approximations by Hayen \textit{et al.} in Refs.~\cite{PhysRevC.99.031301, PhysRevC.100.054323} require a quenched axial coupling constant $g_A=0.9$ fitted by data, while the present \textit{ab initio} calculations use the free-nucleon axial coupling constant $g_A=-1.2701$~\cite{PhysRevD.86.010001} without any empirical adjustment. In the \textit{ab initio} calculations, both the valence-space effective Hamiltonian and effective first-forbidden operator are derived self-consistently in the MBPT framework with chiral EFT forces. We see that both phenomenological and \textit{ab initio} calculations indicate that the self-consistent treatment of first-forbidden transition shape factors is essential to reproduce the experimental bump in the antineutrino spectrum, while the \textit{ab initio} method provides a more fundamental and parameter-free basis for such calculations.

The shadow in Fig.~\ref{U235_spectrum} indicates the theoretical uncertainty arising from many-body calculations. Many-body uncertainties mainly arise from the truncation of the model space, choice of basis parameters (e.g., the HO frequency $\hbar \Omega$), and NO2B approximation. The effect of the NO2B approximation is generally considered to be in the order of 2\% in the medium-mass region~\cite{R.Roth_PhysRevLett.109.052501(2012)}. We have estimated the uncertainty arising from the basis truncation by changing  
$e_{\mathrm{max}}$ and $E_{3\mathrm{max}}$ values, finding that the calculations have been well converged at 
$e_{\mathrm{max} }=14$ and $E_{3\mathrm{max} }=24$. The uncertainty from the basis truncation should be less than 1\%. The newly developed potential of N$^{4}$LO 2NF plus N$^{2}$LO 3NF optimizes
$\hbar\Omega=16~\mathrm{MeV}$ for the $A\sim 100$ mass region~\cite{P.Gysbers_nature.15.428(2019)}. 
We have tested the sensitivity to the $\hbar\Omega$ value by changing $\hbar\Omega$ from 12 to 20 MeV, finding the resulted variation in the spectrum is less than 5\%. Therefore, the total uncertainty arising from many-body calculations should be not larger than 10\%.

\section{\label{sec:level4}SUMMARY}
We have performed the first \textit{ab initio} calculations of first-forbidden $\beta$ transitions, and analyzed their impact on the reactor antineutrino anomaly. Starting from two- plus three-body effective field theory forces, we have derived valence-space effective Hamiltonian and effective operators of forbidden transitions via many-body perturbation theory. Compared with previous works, the present calculations are more self-consistent without the need for empirical quenching factors.

We have computed 20 dominant first-forbidden transitions that contribute significantly to the reactor antineutrino flux and spectrum.
The calculated log$ft$ values are in reasonable agreement with experimental data across $\Delta J=0$, 1, and 2 first-forbidden transitions. However, notable discrepancies between calculations and data remain, especially in the case of $\Delta J=2$ unique first-forbidden transitions. This would suggest the absence of higher-order correlations, such as two-body currents which will be the subject of our future work. Furthermore, calculations reveal that forbidden transitions have non-unity shape factors, which is different from allowed transitions that have a unity shape factor. It is addressed that self-consistent treatment of forbidden-transition shape factors is important in understanding of the anomaly in the reactor antineutrino spectrum. 

The ${ }^{235}$U reactor antineutrino spectrum was focused in the present work. The calculated spectrum with only considering the 20 dominant first-forbidden transitions can reflect the behavior of the experimental spectrum. Note that the dominant first-forbidden transitions govern the reactor antineutrino spectrum in prompt energy higher than 4 MeV. 
The self-consistent non-unity shape factors provide a useful understanding of the ``5 MeV bump'' observed in the ${ }^{235}$U-fission antineutrino spectrum. The downward-sloping shape factors of forbidden transitions redistribute the antineutrino flux toward higher energies, which results in an enhancement in the spectrum.


\section*{\label{sec:level7} Acknowledgments}
We thank B. S. Hu, J. D. Holt and S. Zhang for their useful suggestions and comments. This work has been supported by the National Key Research and Development Program of China under Grants  No. 2024YFA1610900 and No. 2023YFA1606401; the National Natural Science Foundation of China under Grants No. 12335007 and No. 12535008; and the China Postdoctoral Science Foundation under Grant No.\,2024M760489. We acknowledge the High-Performance Computing Platform of Peking University for providing computational resources.

\section*{\label{sec:level7} Data Availability}
The data that support the findings of this article are openly available~\cite{Data}.

\normalem
\bibliography{main}

\end{document}